**Cathodic Carbon Chemically Adsorbs Carbon Dioxide: Why Is it True?**

Vitaly V. Chaban[1] and Nadezhda A. Andreeva[2]

(1) Yerevan State University, Yerevan, 0025, Armenia.

(2) Peter the Great St. Petersburg Polytechnic University, Saint Petersburg, Russia.

**Abstract.** Large-scale applications are waiting for an optimal $CO_2$ scavenger to reinforce CCS and CCU technologies. We herein introduce and succinctly validate a new philosophy of capturing gaseous $CO_2$ by negatively-charged carbonaceous structures. The chemical absorption of $CO_2$ turns out possible thanks to the emergence of significant nucleophilic interaction carbon centers upon applying voltage. The carbonaceous cathode, therefore, may serve as a prototype of a new $CO_2$ sorbent. As a model to simulate chemisorption, we used a small-sized graphene quantum dot (GQD). According to the recorded reaction profiles, the negatively charged GQD containing 16 carbon atoms readily reacts with the $CO_2$ molecule and produces carboxylated GQD. In turn, the activation energy (60 kJ/mol) and energy effect (-55 kJ/mol) for the reaction in water appeared surprisingly competitive in the context of the literature. We hypothesize that the carbonaceous cathode deserves in-depth experimental research as a possible $CO_2$ chemical sorbent. Despite we used GQD for simulations, the encouraging results can be extrapolated to other nanoscale carbons and, more importantly, to the activated carbon species widely employed in modern electrochemical devices.

**Introduction**

Humanity consumes a plethora of energy extracted from various natural sources nowadays. $CO_2$ represents an unavoidable by-product within energy production technologies. In particular, $CO_2$ emits upon the combustion of fossil fuels. Since the concentration of $CO_2$ in the Earth's atmosphere boosts global warming and causes poorly predictable fluctuations in climate, researchers around the world strive to develop capture and valorization routes for $CO_2$.[1-8] The present philosophy in the field assumes that $CO_2$ must be fixed during formation and then either buried (CCS technologies) or valorized (CCU technologies).

The much-discussed peril of global warming and the subsequent interest in $CO_2$, first, as an undesirable acidic gaseous garbage and, second, as a source of carbon atoms have already led to an explosive development of synthetic chemistry. Plenty of organic, inorganic, and hybrid materials are being steadily developed to interact with $CO_2$.[9-15] Physical and chemical sorbents aiming to fix $CO_2$ exhibit very different sets of physicochemical properties and thermodynamics of adsorption. Due to the versatility of the foreseen $CO_2$ scavenging applications, there is no single criterion according to which the elaborated sorbents and CCS technologies can be rated. In the meantime, the successful $CO_2$ sorbents must not cause corrosion of the equipment, must not degrade upon continuous exploitation, and must allow one for energetically cheap recycling.[16-18]

$CO_2$ is primarily a product of combustion, i.e., ultimate oxidation in the oxygen-containing atmosphere. Formed at such conditions, the molecule of $CO_2$ occupies a global energy minimum on the hypothetic potential energy surface plotted for the corresponding atoms. In other words, the lowest energy configuration that can be attained by one carbon atom and two oxygen atoms, excluding nuclear reactions, emerges upon the formation of the $CO_2$ molecule. Such global minimum molecules exhibit limited chemical reactivities due to a limited number of routes to decrease their Gibbs free energies. However, the compound reactions remain possible assuming that their enthalpic gains thanks to a new covalent bond emergence outperform their entropic penalties because of the system's ordering.

One of the most fruitful patterns to bind $CO_2$ thus far has been nitrogen-carbon linking during the carboxamidation of amine derivatives. Apart from various alkyl amines, nitrogen-containing heterocycles demonstrate an articulated affinity to $CO_2$.[19-20] Chemical thinking allows one to predict that certain alternative combinations of the fourteenth and fifteenth groups elements may also represent practical interest. This field remains largely under-explored.

The chemisorption of $CO_2$ that involves carbamate moiety formation appears possible thanks to an electron-rich interaction site at the nitrogen atom of the amino group. The amine-$CO_2$ compound reaction starts with an electrostatic attraction engendered between the electrophilic carbon of $CO_2$ and nucleophilic nitrogen of amine. The covalent bond forms because of the electron-donating properties of nitrogen and the polarization of the carbon atom. It is important to understand that the excess electron density ultimately migrates to the newly formed carboxyl group. Carbamates can exist as zwitterions or conventional molecules depending on the surrounding medium. Carbamates exhibit mediocre kinetic stabilities that result in a certain equilibrium between the reactants (physisorbed $CO_2$) and the product (chemisorbed $CO_2$).

The above-introduced essential knowledge brings us to the idea that the coveted nucleophilicity of the adsorbent can be tuned by various means. More importantly, the capabilities of tuning may ultimately drive chemical designers to even more nucleophilic $CO_2$ scavengers and, therefore, more thermodynamically beneficial $CO_2$ chemisorption reactions.

We herein propose to trial the widely-spread carbonaceous cathodes to chemically fix $CO_2$. We show that the carboxylation reaction occurring at the surface of the electrode is not only possible but energetically favorable upon the application of a certain voltage. We rationalize our unusual observations in terms of the nucleophilicities of the cathode's carbon atoms. As a model of the carbonaceous electrode, we opted to use a tiny graphene quantum dot (GQD) to prove a principle. Since the structures of most chemical structures and nanostructures composed of carbon are fairly similar, the reported results can be readily extrapolated to macroscopic entities.

### Methods and Methodology

We recorded reaction profiles for the $CO_2$ chemisorption by GQD by computing a fraction of the potential energy surface in the framework of hybrid density functional theory (HDFT). The equilibrium electronic structures, charge density distributions, and molecular geometries of reactants, products, and transition states were obtained using the M11 exchange-correlation functional.[21] The molecular wave functions were constructed out of the primitive atomic ones according to the atom-centered double-zeta split-valence polarized 6-31G(d) basis set. The modern HDFT that we opted to employ implicitly includes empirical correction parameters to accurately reproduce the applicable weak long-range attractive (London) forces.[22]

The wave function convergence criterion within the self-consistent field procedure was set to $10^{-7}$ a.u. The rational function optimization algorithm was used to optimize fully restrained and partially restrained geometries along the hypothesized minimum reaction path. The following geometry convergence criteria were chosen: 120 kJ nm$^{-1}$ for the largest force, 80 kJ nm$^{-1}$ for root-mean-squared force, $1.8 \times 10^{-3}$ nm for the largest displacement between subsequent steps, and $1.2 \times 10^{-3}$ nm for root-mean-squared displacement. The convergence of the force-minimization procedure assumed that all four thresholds were fulfilled simultaneously.

The simulated reactions were explored by forcibly moving the chosen reaction coordinate along the desired direction with a uniform step of 0.005 nm. As such, partial geometry optimizations were carried out so that all unrestrained internal coordinates adapt to a single reaction coordinate that evolves. The total potential energies of a series of configurations were plotted against the reaction coordinate. Note that the initial and final states of the simulations represent local minimum stationary points. Saddle points, in turn, represent energy maxima located between the initial and final states. The potential energy difference between the initial state and the saddle point represents the height of the activation barrier. The potential energy difference between the initial state and the deepest minimum state, if any, represents the energetic effect of the reaction. The valid identification of the local maximum was confirmed by determining a single imaginary frequency in the vibrational profile. Note that the potential energy surface scan calculations represent a highly expensive and time-consuming sort of research. The hereby reported results are based on several thousand single-point self-consistent field calculations consequently conducted. All wave functions related to the reaction profiles were solved via unrestricted DFT calculations.

Some simulations included the effects of hydration. Water was represented as an implicit medium according to the polarizable continuum model. The hydration was applied along the entire reaction path because the reactants and the product are expected to exhibit different levels of hydrophilicity.

The chosen molecular configurations were described by their geometrical parameters (covalent and non-covalent distances and angles) and partial electrostatic charges. The latter were obtained according to the Merz-Singh-Kollman algorithm. We use partial atomic charges as a direct measure of the nucleophilicity and electrophilicity of the interaction centers that determine the thermochemistry and activation barriers of the studied reactions.

Gaussian'09D software[23] was called iteratively to conduct electronic-structure calculations. Gaussian'09D is installed in the high-performance supercomputing center TORNADO at the affiliation of the second author. Visual Molecular Dynamics software (ver. 1.9.3)[24] and Gabedit software (ver. 2.5.1)[25] were used to create, edit, and propagate the research problem-relevant Z-matrices, and prepare molecular graphics. The in-home programming developments were used to implement non-standard procedures.

**The Chemisorption by Negatively Charged Carbon**

Figure 1 confirms the impossibility of a direct reaction between $CO_2$ and GQD. Since there is no strong attraction between the carbon atom of $CO_2$ and the carbon atom of GQD the reaction cannot even start. Since there is no excessive electron density within the conduction band of GQD, the possible carboxyl group, as depicted, cannot stabilize. The energy profile recorded along the tested reaction minimum path, therefore, does not contain either a transition state or a minimum corresponding to the $CO_2$ chemisorption product. The plotted potential energies versus distance may be useful for point-by-point comparison with the below-considered case of increased carbon nucleophilicities. At the reaction coordinate, which would otherwise correspond to the formation of the carboxylated product, the potential energy penalty amounts to +300 kJ/mol.

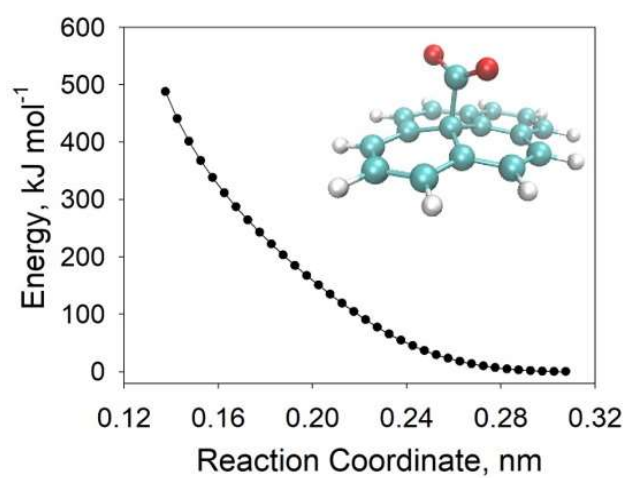

Figure 1. The reaction energy profile corresponds to $CO_2$ reacting with the pristine GQD. The function verifies that no stable products are possible. The selected reaction coordinate is a carbon(GQD)-carbon($CO_2$) distance.

Being a part of the cathode of a certain electric setup, the carbonaceous structure obtains an excessive electron density which is distributed within its conduction band. The separation of charges is intended to create a difference of potentials and, thus, store the electrostatic energy. However, it also changes the chemical properties of the cathode whose adjusted reactivity can be deliberately used to conduct desirable chemical transformations. Figure 2 depicts the reaction profiles in vacuo and water for $CO_2$ capture by a doubly-negatively-charged GQD. The $CO_2$ chemisorption reactions successfully occur in both environments with water playing a substantial stabilizing role for the carboxyl-containing product.

The energetic effect of the $CO_2$ chemisorption by a cathodic GQD in vacuo amounts to +15 kJ/mol, whereas it equals -55 kJ/mol in water. The activation barrier is also smaller in water, 60 kJ/mol, as compared to vacuo, 90 kJ/mol. The saddle points are located at the reaction coordinates of 0.21-0.22 nm indicating the same reaction mechanism in both cases. In turn, the coordinates of the minimum corresponding to the carboxylated product appear close to 0.16 nm. The mentioned distance is within expectations for the carboxyl group emerging on the surface of graphene. Since the valence band of GQD gets additional electrons, e/8 per carbon atom, the nucleophilicities of all carbon atoms belonging to GQD increase. As a result, it becomes possible to conduct the reaction somehow analogous to carboxamidation with fairly competitive thermodynamic parameters. We previously described reminiscent $CO_2$ chemisorptions by phosphonium, sulfonium, and hypothetic ammonium ylides in which cases the carboxylated carbon atom also exhibits the deliberately increased nucleophilicity.[26]

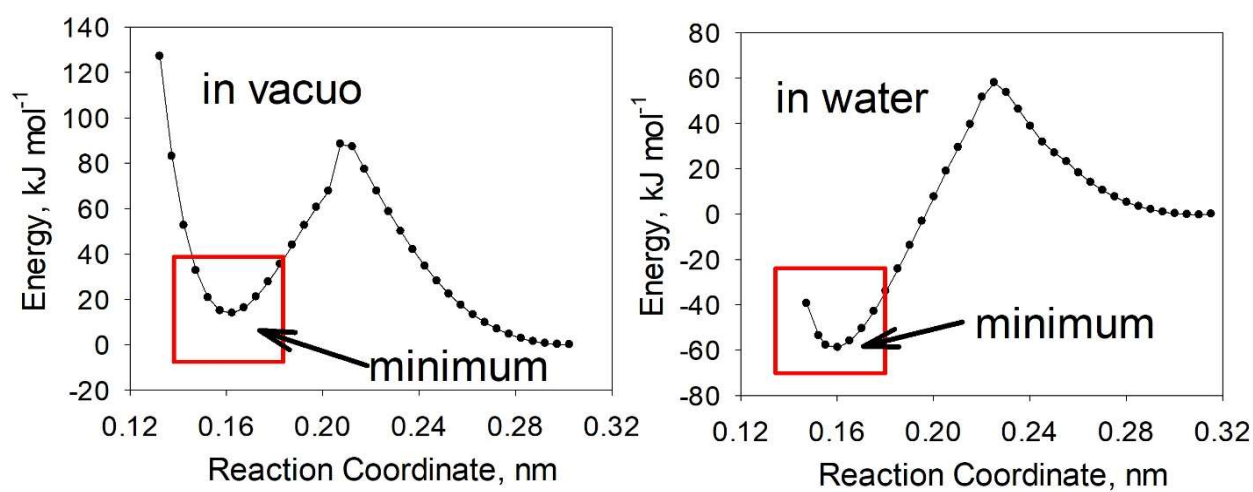

Figure 2. The reaction profiles correspond to $CO_2$ reacting with the doubly-negatively-charged GQD: in vacuo (left) and water (right). The selected reaction coordinate is a carbon(GQD)-carbon($CO_2$) distance.

The increase of electron density within GQD before the $CO_2$ chemisorption initiates is not uniform. Furthermore, not all carbon atoms of the simulated GQD exhibit negative charges. The immediate partial charge depends on the location of the carbon atom and its atomistic environment. For instance, in vacuo, the most nucleophilic sites are rim carbon atoms with partial charges of -0.56e and -0.58e. Other rim carbons have -0.40e and-0.35e, while some of them remain nearly neutral. Some of the central carbons possess charges of -0.08e and -0.002e. In turn, their neighboring carbons are positively charged, +0.21e and +0.27e. Water adds polarization to the charged GQD species. The rim carbon atoms get even more nucleophilic, -0.64e and -0.43e. The central atoms get -0.11e and -0.01e. It is also essential to note that partial charge distributions are highly sensitive to the reaction coordinate and change virtually at every reaction step. Nevertheless, such results are useful to rate the nucleophilicities of carbon as a function of the cathode charge and the stage of the reaction.

**The Chemisorption by Nitrogen-Doped Carbon**

Figure 3 describes three cases of $CO_2$ chemisorption using charged and uncharged nitrogen-doped GQDs. For the sake of generality, we also compare various locations of the alien atoms and their effects on the affinity of the adjacent carbon atoms to $CO_2$.

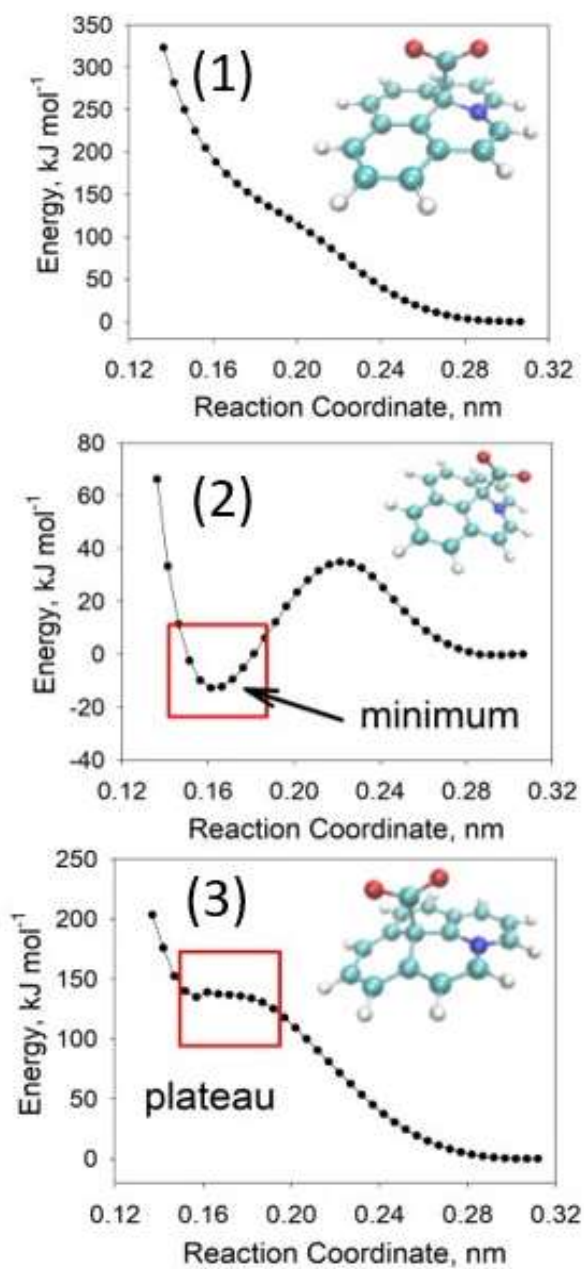

Figure 3. The reaction profiles correspond to $CO_2$ reacting with the nitrogen-doped GQD: (1) uncharged GQD with the nitrogen atom in ortho position to the reaction site; (2) singly-negatively-charged GQD with the nitrogen atom in ortho position to the reaction site; (3) singly-negatively-charged GQD with the nitrogen atom in meta position to the reaction site Note that all hereby presented reactions occur at the carbon sites of the N-doped GQDs. The selected reaction coordinate is a carbon(GQD)-carbon($CO_2$) distance.

The non-charged N-doped graphene exhibits no affinity to $CO_2$. The energetic penalty is, however, noticeably smaller than that in the case of pristine graphene, +200 vs.

+300 kJ/mol at the reaction coordinate of the possible carboxylated GQD. One concludes that the nitrogen-doping effect is not commensurate with the effect of the cathode charging. Note that we deliberately discuss only the carbon site carboxylation in the present work, whereas carboxamidation at the nitrogen dopants would be a more energetically feasible route.

The charged cathode bearing e/16 excessive electron density per heavy atom exhibits quite a competitive chemisorption energetics. The barrier height amounts to 35 kJ/mol. In turn, the energy effect is -15 kJ/mol. However, the above result is only relevant to the ortho positions of the reacting sites relative to the alien nitrogen. The meta position fosters an energy plateau that is unlikely enough to provide a thermodynamically and kinetically stable chemisorption product. Further exploration of the possibilities that N-doped nanoscale carbon offers may lead to useful fundamental generalizations and open novel avenues for robust applications.

### Conclusions and Final Remarks

We hereby introduced the idea of $CO_2$ sorption by the negatively charges pristine species of carbon, such as graphite, graphene, nanotubes, activated carbon, etc. Upon application of voltage, the cathodic carbon gains necessary nucleophilicities and reacts with $CO_2$ to give rise to the carboxylated GQD. The activation barrier of such chemisorption appears surprisingly low, whereas the energetic effect is more favorable as compared to alkyl amines. The hydration of the system stabilizes the carboxyl-containing product since it is more polar than the reactants.

Our electronic-structure HDFT simulations univocally prove the possibility of capturing $CO_2$ at the cathode containing nanoscale or macroscale carbons. Future research must investigate the impact of the graphene sheet size on the energetics of chemisorption. Larger sheets must be simulated to consider the carboxylation of the carbon atoms located farther from the sheet edges. Moreover, it is interesting to consider the carboxylation of the rim carbon atoms since the latter were found to attain the strongest nucleophilicities. By systematically alternating the total charge of the cathode and the number of its atoms, it would be possible to evaluate the balanced parameters of the $CO_2$ scavenger to both obtain high $CO_2$ capacities and preserve the electrode's stability. Due to the non-uniform distribution of the excessive electron density over the charged carbonaceous cathode, it will be important to search for a minimum reaction path as a function of the target destination of the captured $CO_2$ molecule. The kinetics of the sorbent-sorbate system after the removal of voltage remains to be described.

We, furthermore, attempted to emulate the increased nucleophilicities of the scavenger's carbon atoms by investigating nitrogen-doped graphene. We found that nitrogen atoms in the vicinity of the reacting carbon atoms boost chemisorption at twice as smaller electrode's negative charge. Herewith, nitrogen in the ortho position relative to the carboxylated site exhibits much better performance as compared to the one in the meta position.

The provided proof of principle holds promise to open a vibrant avenue of novel solutions to mitigate atmospheric concentrations and valorize $CO_2$. The evaluated activation barriers and reaction effects urge experimental verification by measuring the direct $CO_2$ capacities of the cathodic macroscopic and nanoscopic carbons.

## Acknowledgments

Some of the reported results were obtained using computational resources of Peter the Great Saint-Petersburg Polytechnic University Supercomputing Center (www.spbstu.ru).

## Authors

Prof. Vitaly V. Chaban is widely and internationally referenced in the fields of molecular design, physical chemistry, and solution chemistry having authored over ten books and two hundred research papers. He developed several important methods to gain practically-important insights from in-silico calculations. Dr. Nadezhda A. Andreeva is a researcher and educator in physico-mathematical sciences with primary expertise in carbon dioxide capture technologies and proactive-integral teaching of high school and university students.

## Contributions of Authors

V.V.C. invented the idea; created the research program; developed the programming instrumentation; wrote the manuscript. N.A.A. conducted the electronic-structure simulations under supervision; analyzed the obtained data; plotted results. Both authors participated in critical scientific discussions and shared their interpretations to generate new chemical knowledge.

**Conflict of interest**

The authors hereby declare no financial interests that might have biased our philosophy, design of in-silico experiments, and interpretations of the obtained results.

**Author for correspondence**

Inquiries regarding the scientific content of this paper including collaboration requests shall be directed through electric mail to Prof. Dr. Vitaly V. Chaban at vvchaban@gmail.com.

**References**

1.      Aquino, A. S.; Bernard, F. L.; Borges, J. V.; Mafra, L.; Dalla Vecchia, F.; Vieira, M. O.; Ligabue, R.; Seferin, M.; Chaban, V. V.; Cabrita, E. J.; Einloft, S., Rationalizing the role of the anion in CO2 capture and conversion using imidazolium-based ionic liquid modified mesoporous silica. *RSC Adv.* **2015,** *5* (79), 64220-64227.

2.      Bernard, F. L.; Dalla Vecchia, F.; Rojas, M. F.; Ligabue, R.; Vieira, M. O.; Costa, E. M.; Chaban, V. V.; Einloft, S., Anticorrosion Protection by Amine-Ionic Liquid Mixtures: Experiments and Simulations. *J Chem Eng Data* **2016,** *61* (5), 1803-1810.

3.      Bernard, F. L.; Polesso, B. B.; Cobalchini, F. W.; Donato, A. J.; Seferin, M.; Ligabue, R.; Chaban, V. V.; do Nascimento, J. F.; Dalla Vecchia, F.; Einloft, S., CO2 capture: Tuning cation-anion interaction in urethane based poly(ionic liquids). *Polymer* **2016,** *102*, 199-208.

4.      Bernard, F. L.; Rodrigues, D. M.; Polesso, B. B.; Donato, A. J.; Seferin, M.; Chaban, V. V.; Vecchia, F. D.; Einloft, S., New cellulose based ionic compounds as low-cost sorbents for CO2 capture. *Fuel Process Technol* **2016,** *149*, 131-138.

5.      Baldissera, A. F.; Schütz, M. K.; dos Santos, L. M.; Vecchia, F. D.; Seferin, M.; Ligabue, R.; Costa, E. M.; Chaban, V. V.; Menezes, S. C.; Einloft, S., Epoxy resin-cement paste composite for wellbores: Evaluation of chemical degradation fostered carbon dioxide. *Greenh. Gases Sci. Technol.* **2017,** *7* (6), 1065-1079.

6.      Bernard, F. L.; Polesso, B. B.; Cobalchini, F. W.; Chaban, V. V.; Do Nascimento, J. F.; Dalla Vecchia, F.; Einloft, S., Hybrid Alkoxysilane-Functionalized Urethane-Imide-Based Poly(ionic liquids) as a New Platform for Carbon Dioxide Capture. *Energy Fuels* **2017,** *31* (9), 9840-9849.

7.    Vieira, M. O.; Monteiro, W. F.; Ligabue, R.; Seferin, M.; Chaban, V. V.; Andreeva, N. A.; do Nascimento, J. F.; Einloft, S., Ionic liquids composed of linear amphiphilic anions: Synthesis, physicochemical characterization, hydrophilicity and interaction with carbon dioxide. *J Mol Liq* **2017,** *241*, 64-73.

8.    Bernard, F. L.; Duczinski, R. B.; Rojas, M. F.; Fialho, M. C. C.; Carreño, L. Á.; Chaban, V. V.; Vecchia, F. D.; Einloft, S., Cellulose based poly(ionic liquids): Tuning cation-anion interaction to improve carbon dioxide sorption. *Fuel* **2018,** *211*, 76-86.

9.    Duczinski, R.; Bernard, F.; Rojas, M.; Duarte, E.; Chaban, V.; Vecchia, F. D.; Menezes, S.; Einloft, S., Waste derived MCMRH- supported IL for CO2/CH4 separation. *J. Nat. Gas Sci. Eng.* **2018,** *54*, 54-64.

10.    Vieira, M. O.; Monteiro, W. F.; Neto, B. S.; Ligabue, R.; Chaban, V. V.; Einloft, S., Surface Active Ionic Liquids as Catalyst for CO2 Conversion to Propylene Carbonate. *Catal Lett* **2018,** *148* (1), 108-118.

11.    Bernard, F. L.; Rodrigues, D. M.; Polesso, B. B.; Chaban, V. V.; Seferin, M.; Vecchia, F. D.; Einloft, S., Development of inexpensive cellulosebased sorbents for carbon dioxide. *Brazil J Chem Eng* **2019,** *36* (1), 511-521.

12.    Campbell, S.; Bernard, F. L.; Rodrigues, D. M.; Rojas, M. F.; Carreño, L.; Chaban, V. V.; Einloft, S., Performance of metal-functionalized rice husk cellulose for CO2 sorption and CO2/N2 separation. *Fuel* **2019,** *239*, 737-746.

13.    Vieira, M. O.; Monteiro, W. F.; Neto, B. S.; Chaban, V. V.; Ligabue, R.; Einloft, S., Chemical fixation of CO 2 : the influence of linear amphiphilic anions on surface active ionic liquids (SAILs) as catalysts for synthesis of cyclic carbonates under solvent-free conditions. *React. Kinet. Mech. Catal.* **2019,** *126* (2), 987-1001.

14.    Faria, D. J.; dos Santos, L. M.; Bernard, F. L.; Pinto, I. S.; Romero, I. P.; Chaban, V. V.; Einloft, S., Performance of supported metal catalysts in the dimethyl carbonate production by direct synthesis using CO2 and methanol. *J. CO2 Util.* **2021,** *53*.

15.    Nisar, M.; Bernard, F. L.; Duarte, E.; Chaban, V. V.; Einloft, S., New polysulfone microcapsules containing metal oxides and ([BMIM][NTf2]) ionic liquid for CO2capture. *J. Environ. Chem. Eng.* **2021,** *9* (1).

16.    Andreeva, N. A.; Chaban, V. V., Electronic and thermodynamic properties of the amino- and carboxamido-functionalized C-60-based fullerenes: Towards non-volatile carbon dioxide scavengers. *J Chem Thermodyn* **2018,** *116*, 1-6.

17.    Chaban, V. V.; Andreeva, N. A., *Combating Global Warming: Moderating Carbon Dioxide Concentration in the Earth's Atmosphere through Robust Design of Novel Scavengers*. LAP LAMBERT Academic Publishing: 2018; p 164.

18.    Chaban, V. V.; Andreeva, N. A.; Vorontsov-Velyaminov, P. N., A Weakly Coordinating Anion Substantially Enhances Carbon Dioxide Fixation by Calcium and Barium Salts. *Energy & Fuels* **2017,** *31* (9), 9668-9674.

19.    Chaban, V. V.; Andreeva, N. A., Amination of Five Families of Room-Temperature Ionic Liquids: Computational Thermodynamics and Vibrational Spectroscopy. *Journal of Chemical & Engineering Data* **2016,** *61* (5), 1917-1923.

20.    Andreeva, N. A.; Chaban, V. V., Amino-functionalized ionic liquids as carbon dioxide scavengers. Ab initio thermodynamics for chemisorption. *J Chem Thermodyn* **2016,** *103*, 1-6.

21.    Peverati, R.; Truhlar, D. G., Performance of the M11 and M11-L density functionals for calculations of electronic excitation energies by adiabatic time-dependent density functional theory. *Physical Chemistry Chemical Physics* **2012,** *14* (32), 11363-11370.

22.     Risthaus, T.; Grimme, S., Benchmarking of London Dispersion-Accounting Density Functional Theory Methods on Very Large Molecular Complexes. *Journal of Chemical Theory and Computation* **2013,** *9* (3), 1580-1591.

23.     Frisch, M. J.; Trucks, G. W.; Schlegel, H. B.; Scuseria, G. E.; Robb, M. A.; Cheeseman, J. R.; Scalmani, G.; Barone, V.; Petersson, G. A.; Nakatsuji, H.; Li, X.; Caricato, M.; Marenich, A. V.; Bloino, J.; Janesko, B. G.; Gomperts, R.; Mennucci, B.; Hratchian, H. P.; Ortiz, J. V.; Izmaylov, A. F.; Sonnenberg, J. L.; Williams; Ding, F.; Lipparini, F.; Egidi, F.; Goings, J.; Peng, B.; Petrone, A.; Henderson, T.; Ranasinghe, D.; Zakrzewski, V. G.; Gao, J.; Rega, N.; Zheng, G.; Liang, W.; Hada, M.; Ehara, M.; Toyota, K.; Fukuda, R.; Hasegawa, J.; Ishida, M.; Nakajima, T.; Honda, Y.; Kitao, O.; Nakai, H.; Vreven, T.; Throssell, K.; Montgomery Jr., J. A.; Peralta, J. E.; Ogliaro, F.; Bearpark, M. J.; Heyd, J. J.; Brothers, E. N.; Kudin, K. N.; Staroverov, V. N.; Keith, T. A.; Kobayashi, R.; Normand, J.; Raghavachari, K.; Rendell, A. P.; Burant, J. C.; Iyengar, S. S.; Tomasi, J.; Cossi, M.; Millam, J. M.; Klene, M.; Adamo, C.; Cammi, R.; Ochterski, J. W.; Martin, R. L.; Morokuma, K.; Farkas, O.; Foresman, J. B.; Fox, D. J. *Gaussian 09*, Gaussian, Inc.: Wallingford CT, 2009.

24.     Humphrey, W.; Dalke, A.; Schulten, K., VMD: Visual molecular dynamics. *Journal of Molecular Graphics & Modelling* **1996,** *14* (1), 33-38.

25.     Allouche, A. R., Gabedit-A Graphical User Interface for Computational Chemistry Softwares. *Journal of Computational Chemistry* **2011,** *32* (1), 174-182.

26.     Chaban, V. V.; Andreeva, N. A.; Voroshylova, I. V., Ammonium-, phosphonium- and sulfonium-based 2-cyanopyrrolidine ionic liquids for carbon dioxide fixation. *Phys Chem Chem Phys* **2022,** *24* (16), 9659-9672.